# On the nature of the in-ecliptic interplanetary magnetic field's two-humped distribution at 1AU

O. Khabarova · V. Obridko

*Heliophysical Laboratory, Institute of Terrestrial Magnetism, Ionosphere and Radio Wave Propagation RAS (IZMIRAN), Troitsk, Moscow Region, 142190 Russia*

Phone: +79037807709

Fax: +74953331248

E-mail: olik3110@aol.com

**Abstract**

It was found out that the distribution's shape of the in-ecliptic (as well as radial) component of the interplanetary magnetic field (IMF) significantly changes with the heliocentric distance, which poorly corresponds to classical models of the solar wind and the interplanetary magnetic field (IMF) expansion. For example, distributions of the radial photospheric and the source surface's magnetic field in the ecliptic plane are Gaussian-like, the distribution of the radial IMF component Br at the Earth orbit demonstrates two-humped shape, and it becomes again Gaussian-like at 3-4 AU. These differences lead to lack of correspondence between simulations of the IMF behaviour at 1 AU and observations. Our results indicate that picture of the IMF expansion into space is more complicated than usually considered, and the sector structure is not the only source of the two-humped shape of the in-ecliptic (or Br) IMF component. We have analysed data from different spacecraft at the distances from 0.29 AU to 4 AU and found that the shape of the radial IMF component distribution strongly depends on a heliocentric distance and a heliolatitude. The "two-humped IMF" effect is most brightly expressed at low heliolatitudes at 0.7-2 AU, but it fully disappears at 3-4 AU. There is also dependence of the IMF distributions' view on a solar cycle due to active processes, such as solar flares and CMEs. We suppose that the in-ecliptic solar wind field at 1 AU is influenced by solar active regions in a high degree, and actually its distribution is the three-humped: two humps reflect the impact of flows from the middle and high heliolatitudes and the third one is the theoretically expected distribution from the solar field nearby the heliomagnetic equator. Vanishing of the IMF zero-component with the distance from the Sun partially could be a result of a magnetic reconnection at the current sheets in the solar wind.

---

# 1. Introduction

### 1.1. What is wrong with the in-ecliptic IMF distribution?

Using OMNI2 database, it is easy to find that distributions of the Bx and By horizontal components of the interplanetary magnetic field (IMF) in GSE coordinate system (or the radial IMF component Br in RTN coordinate system) at the Earth orbit have a striking hole in the area ±1 nT around zero strength component. There are two peaks of the distributions: one negative and one positive, while the vertical Bz component is normally distributed around zero value. This effect has been considered as well-known since the beginning of the space era (see, for instance, Veselovsky, Dmitriev, and Suvorova, 2010). Indeed, for many years a two-humped distribution of the in-ecliptic IMF at 1 AU has been simply explained by geometrical considerations. It is supposed that sector structure is responsible for the effect (Russell, 2000). Objectively, the heliospheric current sheet (HCS), where nil magnetic lines always occur, is rather thin, and it passes the Earth rather fast (time of the Earth's stay in a positive or negative sector is much longer). Therefore we could expect to observe a lot of positive and negative in-ecliptic IMF strength values in comparison with a small number of the IMF strength zeros practically at any heliocentric distance.

This explanation seemed so obvious and logical that the effect practically had not been discussed in the literature so long as Belov, Obridko, and Shelting (2006) denoted the fact that IMF in-ecliptic distribution at 1 AU looks absolutely different with the histogram of the radial solar magnetic field along the ecliptic plane (according to measurements of solar magnetometers). If we trust the most popular MHD coronal expansion models from Biermann (1957), Parker (1958),

Pneumann and Kopp (1971) to nowadays (Schwadron and McComas, 2005), the distribution of the magnetic field in photosphere or at the source surface along the ecliptic plane section of the Sun should has the similar view with the in-ecliptic IMF at 1 AU. It is expected that the projection of the Earth onto the Sun crosses the warped heliomagnetic equator approximately as many times as the HCS crosses the Earth orbit at 1 AU. As a result, we should see the one-type distributions of the corresponded magnetic fields, unchanging with the heliocentric distance. So the results by Belov, Obridko, and Shelting, 2006 indicate some poor-understood points in the picture of the IMF temporal and spatial distribution. Modern solar magnetometers measure a radial component of the magnetic field of the Sun with an accuracy and time resolution good enough for the solar field distribution comparison with the near-Earth IMF strength distribution obtained from spacecraft data, so the discussed phenomenon can not be explained by technical problems.

Hypothetically, the complicity of the interplanetary magnetic filed expansion (like a tendency of the near-equator corona to expand along the HCS, the IMF entangle, "ballerina skirt" effect, and the HCS motion) could lead to more frequent IMF zeros occurrence at 1 AU in comparison with number of zeros, measured by solar magnetometers, but in reality we see the opposite.

The main questions of the current investigation are: How can it be? How does the in-ecliptic IMF vary with the distance from the Sun? What is the nature of the "two-humped IMF" phenomenon?

**1.2. Consequent problems of modelling**

An additional indication of some uncertainty about the rules of photospheric field extension is that the prognostic coronal expansion models give low values of the in-ecliptic IMF strength more often than they are really observed at 1 AU. (Belov, Obridko, and Shelting, 2006). One example is given in Figure 1. $B_L$ is the in-ecliptic IMF component at 1 AU. Black line with daggers is the result of the modelling, and black points are the observed values of $B_L$. It is easy to see that the daggers' number essentially exceeds the number of experimentally observed points in the selected area around zero value. This is obviously a result of taking of Haussian-like distribution at the source surface for calculations.

## 1.3. Two humps of the in-ecliptic IMF distribution give "a floor" of the IMF strength

Observation of the two-humped distribution of the in-ecliptic IMF is the primary cause of the shift of the module IMF distribution's peak from zero value, i.e. the IMF strength has a "floor" (some lack of low values). Figure 1 is an evidence of such "insufficiency" of near-zero IMF strength values. Svalgaard and Cliver (2007) provided the other evidence for a "floor" of the solar wind magnetic field. They noticed that a minimum value of the yearly averaged in-ecliptic IMF strength is rather high; it is 4.6 nT for IMF measurements at 1 AU from 1965 to present.

There are several possible explanations of the observed in-ecliptic IMF zero depression behind the commonly accepted view on sector structure to be the main cause of the effect:

1. Original radially-directed photospheric magnetic field, measured by solar magnetometers, significantly bends in the vertical direction with heliocentric distance, so most of the former in-ecliptic IMF zeros are included into Bz (vertical) IMF component at 1 AU. Strong inclination of the HCS to the ecliptic plane at 1 AU could be a cause of this effect.

2. The solar magnetic field expands not radially, so streams from high and middle latitudes mix with streams coming from the areas near the heliomagnetic equator. As a result, we observe practically unexpected magnetic field quite different from the one calculated for 1 AU on the base of models, using solar filed at the source surface as an initial condition.

3. Some insufficiency of near-zero values in the IMF horizontal components at 1 AU in comparison with the source surface field could indicate vanishing of in-ecliptic IMF zeros somewhere on the way from the Sun to the Earth. So, it is possible to suppose that IMF neutral lines partially disappear with heliocentric distance due to some nonlinear processes in the solar wind plasma.

4. A combination of the three mentioned-above causes is quite possible.

In this work we intend to investigate this effect and find the most convincing explanation of the specific shape of in-ecliptic IMF distribution, observed at 1 AU.

# 2. Features of the in-ecliptic magnetic field distribution: experimental facts

We believe that experimental facts are always primary, so we will start with the discussion: how features of the IMF strength distribution do vary with the solar cycle, the heliocentric distance, and heliomagnetic latitude. For the best understanding of the picture of the IMF temporal and spatial distribution in space we have used OMNI2 daily data as well as data of different spacecraft (*Helios 2*, *Pioneer Venus Orbiter*, *Ulysses* and *Voyager 1*) from 1977 to 2009. Certainly, no one spacecraft provides data for the all time range, but number of measurements is enough to draw statistically proved conclusions. All possible explanations and comparisons of the obtained results with theories and hypotheses will be given in the next section.

## 2.1 DISTRIBUTIONS AT 1 AU

First of all, let's look closer at the difference between shapes of the source surface magnetic field histogram and the horizontal IMF components distribution at 1 AU ($B_x$, and $B_y$ in GSE coordinate system), mentioned in the Introduction. The source surface is traditionally supposed to be at 2.5 solar radii.

Year by year $B_x$ and $B_y$ distributions demonstrate insufficiency of zero strength component in comparison with the source surface radial magnetic field distribution ($B_r$), which is Gaussian-like (Figure 2ab). At the same time, the vertical IMF component $B_z$ is sharply distributed around zero value (Figure 1b). IMF data used for mapping are from OMNI2 hourly database, and $B_r$ is calculated along the line of the Sun's section by the ecliptic plane on the base of solar synoptic charts (Wilcox Solar Observatory data).

The horizontal IMF components have a wide distributions, in contrast to $B_z$. This is a pure geometrical effect, as the IMF at 1 AU predominantly lies close to the ecliptic plane, so the vertical field component mostly equals zero with small deviations from it due to the influence of structures such as coronal mass ejections (CMEs) or corotating interaction regions (CIRs).

The observed difference between $B_x$, $B_y$ and $B_r$ histograms looks very intriguing. Let's investigate how the histograms change with a solar cycle.

## 2.2 DEPENDENCE ON THE SOLAR CYCLE

Analysis of the IMF distribution changes over the solar cycle was performed for 1977-2009. We suppose the intervals of 1979-1981, 1989-1991, and 1999-2001 to be solar maxima, and 1977, 1985-1986, 1996-1997, 2007-2009 to be minima (according to Figure 3). Histograms of Bx, By and Bz IMF components and of the source surface magnetic field Br are shown in Figure 4 for these periods.

The histograms look very different at solar maximum and minimum. The first feature is their falling down and spreading during solar maximum. Such a behaviour of Bx, By and Bz histograms (Figure 4a,b,c) can be explained by highly disturbed solar wind plasma by increased number of CMEs at solar maximum, which bring strong magnetic fields in their body.

The histogram of the vertical Bz IMF component (Figure 4c) falls at solar maximum together with Bx and By. This fact allows rejecting the supposition about "zeros' spilling over from Bx and By into Bz" (see the Introduction). As we know, Parker's solar wind classical model includes only horizontal components of the interplanetary magnetic field. The nature of the vertical one is still debated; it possibly appears as a result of non-linear processes in the real solar wind plasma on the way from the Sun to the Earth. If that is true, Bz could have the histogram shape, which does not depend on behaviour of the in-ecliptic IMF. For example, it could increase with a distance or vary during solar cycle in the way different from the Bx and By IMF components. Unity of Bx, By and Bz changes with the solar cycle may be considered as a sign of initially existed vertical component of the solar magnetic field, expanding into space, so place of the Bz origination is not the solar wind.

Source surface radial magnetic field Br (Figure 4d) also has a tendency to depression and spreading during solar maxima due to the change of quasi-dipole solar field at minimum for multipole field at maximum. Deep fall of the Br histogram says that at solar maximum all values of Br are practically equiprobable, magnetic field is very strong even in near-equator latitudes, and the ecliptic plane crosses the heliomagnetic equator relatively seldom (because the equator's shape is very far from the straight line in this period, it is waved, according to the number of multipoles).

It is remarkable that the IMF components drop not in such a degree. So, behaviour of the IMF at 1 AU is most possibly not highly influenced by original solar field, measured along the ecliptic plane projection onto the Sun.

It is important to mention that the some asymmetry of the horizontal IMF components histograms' shape in Figure 4ab during solar minimum is a result of dominating sign of the active regions. If we take longer time period into consideration, this effect will disappear and the histograms become symmetrical.

## 2.2 DEPENDENCE ON A RADIAL DISTANCE FROM THE SUN

We considered above the distribution of the radial magnetic field of the Sun at the solar wind source surface Br, and the IMF components distribution at 1 AU. Here we will trace changes in the Br histogram with heliocentric distance. Taking into account that the distribution of the radial IMF component Br practically coincides with the Bx distribution at distances up to 4 AU considered below (except for sign), we will consider only Br IMF component for our statistical analysis.

Figure 5ab shows the Br distribution at the photosphere and the source surface for 1977-2009. Both fields represented in Figure 4ab are quasy-normally distributed. The next Br histogram (Figure 5c) is based on the *Pioneer Venus Orbiter* daily data from 5 December, 1978 to 31 December, 1992. The Venus heliocentric distance is 0.7 AU. The effect of two-humped distributions is expressed here in a high degree. Investigations show that 0.7 AU in-ecliptic IMF Br depends on the solar cycle like the in-ecliptic magnetic field at 1 AU do (see Figure 4).

Figure 5d represents Br histograms from *Voyager 1* that measured the IMF practically in ecliptic plane for the distances, which we will analyse below. As *Voyager 1* passed the distance of 2-4 AU rapidly, for the best statistics we calculated the histograms based on the not daily, but the hourly data. The Br histogram at 2.00-2.99 AU (from 12:00 8 January, 1978 to 11:00 24 April, 1978) still demonstrates weak "two-humped" effect, but the distribution at 3.00-3.99 AU (from 12:00 24 April, 1978 to 11:00 24 August, 1978) has no any fall around Br zero value. Therefore, the "two-humped" distribution phenomenon appears in ecliptic plane somewhere before 0.7 AU and then get vanish with a distance.

More precise picture of the Br distribution transformation with a heliocentric distance can be obtained from *Helios 2* mission data. *Helios 2* had a very elongated orbit and was a unique space probe that approached to the Sun closer than the Mercury, at a distance of 0.29 AU. We tested Br at four distance intervals (as shown in Figure 6) for all available daily data between 16 January, 1976 and 5 March, 1980. It is interesting that there are not two, but three humps of the Br distribution up to 0.8 AU. We will keep in mind and discuss the nature of this phenomenon later.

Thus, the shape of Br distribution is far from the expected Gaussian even at small heliocentric distances.

## 2.3 DEPENDENCE ON HELIOSPHERIC LATITUDE

The next question is about changes in the Br histogram with heliospheric latitudes. Let's imagine that we know nothing about the rules of the IMF expansion into space, i.e. we can suppose it to be not radial, but also having non-radial deflection. Signs of that and possible models are reported frequently (see, for instance, Smith and Balogh, 1995; Fisk, 1996). Hence it is possible to see input of the high-latitude IMF into the in-ecliptic IMF distribution.

As we mentioned in the Introduction, Svalgaard and Cliver (2007) pointed out a very important IMF feature - its "floor" in the solar wind. The "floor" is a baseline state, which the IMF falls to, at the sunspot number approaching to zero. They experimentally estimated the floor as ~4.6 nT at 1 AU in the ecliptic plane, and as ~3.2 nT at high heliolatitudes (upper than 37°). From this point of view, the IMF features should be different at different heliospheric latitudes. If we suppose that the floor is partially determined by the "two-humped" effect (see the Introduction), the radial component Br should has different distributions at low and high latitudes. On the other hand, there is a strong tendency to consider the radial IMF Br to be independent of heliolatitude and a solar cycle (Smith and Balogh, 1995; Smith and Balogh, 2008; Loockwood *et al*., 2004 ). So, it is necessary to look at features of the latitudinal Br distribution closely.

We divided *Ulysses* spacecraft hourly data on categories, according to heliocentric distance and heliolatitudes. High and low latitudes were considered to be above or below 40° respectively. Figure 7 shows the Br distribution at high (Figure 7a) and low (Figure 7b) latitudes at distances from 1AU to 4AU during the time interval from 1990 to 2008. Each distribution includes data for several years, corresponding to different solar cycle phases, so we see here some averaged picture (like in Figure 2). For example, 1-2 AU distribution in Figure 7a was mapped on the base of 1994-1995, 2001, 2007-2008 data, and the same one in Figure 7b represents the Br histogram for 1990-91, 1995, 2001, 2007. Number of hours was, respectively, 11996 and 10505.

Figure 7a gives us possibility to imagine a picture of the possible Br superposition at distances 1-4 AU from the Sun under condition of the IMF deflection (distributions for the both South and North high-latitude Br were mixed in one in Figure 7a). The high-latitude Br component still has a "two-humped" distribution at 3-4 AU, when the low-latitude Br is already one-humped.

The Br distributions in Figures 7a and 7b have different view. High-latitude Br distributions have no long tails of high values at any distances in comparison with Figure 7b. For example, at 1-2 AU the "humps" maxima correspond to ±0.8-1.0 nT at high latitudes, and to ±1.5-1.6 nT at low latitudes. The tail of the 1-2 AU distribution in Figure 7a is limited by values ~ ±2nT, but it reaches the values more than ±4 nT in Figure 7b.

More precisely this difference can be viewed in Figure 8, where the Br distributions both at high and low latitudes are given for 2007 without "superposition" at high latitudes. The positive and negative humps' maxima are observed practically at the same values both for the low-latitude and high latitude Br, but the difference of the tails is essential (the low-latitude Br is distributed significantly widely than the high-latitude one).

We can conclude that Br is weaker at high heliolatitudes than at low latitudes, if we do not recalculate it to the Earth orbit and do not average it over several solar cycles in the way by, for example, Smith and Balogh (1995). If we take into account that any recalculations could be based on wrong assumptions, it is possible to suppose that weak high-latitude Br could give stable input in the in-ecliptic Br "humps".

The solar cycle effect exists both for Br changes at low and high heliolatitudes (not shown), but it is not so clear as for a long time interval used in Figure 4, so we can conclude nothing definite about the difference between the effect in polar and in-ecliptic IMF at different phases of the solar cycle.

## 3. Possible causes of the effect

Zero-component of the magnetic field at the discussed histograms could appear, first of all, as the result of real spacecraft measurements, and then as a result of averaging. If we take, for example, 1-day data and choose a rather wide histogram span, we actually obtain zero-value due to calculation of the mean value of precisely measured positive and negative values around zero strength for near-zero span. So, when we speak about zero insufficiency (or about the "two-humped" in-ecliptic IMF), we mean existence of the effect for daily data: an artificial "averaging-made" near-zero hump could be seen in histograms based on daily data (for example, as in Figure 6), but the histograms' fall around zero value definitely is not an artefact.

Summarizing the results obtained above, we conclude that observed picture of the IMF distribution is rather far from the one predicted from the main models of the solar wind expansion. The "two-humped" effect is not observed in the photosphere or at the solar wind source surface in the ecliptic plane, but it is detectable at 0.29 AU. It can be easily seen at 0.8-1 AU, and still remains observable at 2-3 AU at low heliolatitudes. Hereafter we will give an our explanation of these features of the IMF distribution.

### 3.1 THE HELIOSPERIC CURRENT SHEET PLAYS A MINOR ROLE

As we have already mentioned, the idea that the heliospheric current sheet (HCS) is the only cause of the in-ecliptic IMF "zero insufficiency", should be rejected due to the fact that the ecliptic plane is expected to cross the HCS as frequently (or even more often) as it crosses the heliomagnetic equator, so zero-lines' occurrence must be the same at different heliocentric distances (at least, the in-ecliptic IMF must have the similar distribution shape at the Sun and in the solar wind).

Meanwhile, we can make one more assumption: if majority of zeros in the IMF distribution are associated with the HCS, its strong inclination (its turn in the vertical direction with a heliocentric distance) might result to zeros' "spilling over" from the horizontal IMF components into the vertical one. Let's analyze the HCS properties: its inclination to the ecliptic plane and its input into the "two-humped distribution" effect. We will show here why the sector pattern can not be only cause of the two-humped distribution of the in-ecliptic IMF at 1 AU.

The generally accepted opinion about the IMF and the HCS geometry is illustrated in Figure 9. Since Wilcox and Ness (1965) discovered the heliospheric current sheet, it has been considered as the heliomagentic equator extension into the heliosphere, i.e. as a surface that separates opposite magnetic polarity. The average position of the solar magnetic equator is tilted relative to the ecliptic plane (Figure 9a), and a sector structure of the IMF is observed at the Earth orbit.

Besides, some problems emerge when we try to predict the HCS position, thickness, and inclination towards the ecliptic plane. First of all, the heliomagnetic equator is usually warped (especially at solar maxima periods) due to multipole structure of the solar field, so the Earth passes through regions of different polarity more frequently and the HCS inclination changes more unpredictable than expected.

According to the classical models, the IMF twists into a Parker Spiral with angle α=45° at the Earth orbit (see Figure 7b). In reality α is distributed widely, mainly from -55 to -31 (Veselovsky, Dmitriev, and Suvorova, 2010). Additionally, a Parker Spiral is waved as a "ballerina skirt" due to existence of the angle ~7° between the solar magnetic and rotation axes. Actually, deviations from the Parker spiral direction are observed very often. The highest disturbed IMF profile is observed at solar maxima, when the HCS looks like a ''conch shell'' (Riley, Linker, and Mikic, 2002). The nature of this effect has been discussed many times, but presently, there is not a common agreement on this despite the existence of several models trying to explain and describe the observed IMF picture (Schwadron and McComas, 2005; Riley and Gosling, 2007).

The observed HCS inclinations poorly correspond with predicted ones. For example, Ho *et al.* (1997) used the Stanford source surface magnetic field model for prediction of the HCS crossing locations at 1 AU and 1.4 AU. They found differences between calculations of the location of the neutral line and real data up to 25 degrees in longitude. It was found out that the best way for prediction of its position and inclination is not the MHD models' or the current sheet model's use, but use of high-resolution Michelson Doppler Imager synoptic charts or computations, based on relatively simple potential field source surface (PFSS) model, where the photospheric magnetic field is assumed to be radial everywhere on the photosphere (Neugebauer *et al.*, 1998; Burton, *et al.*, 1994; Zhao *et al.*, 2006).

Summarizing, we conclude that the classical model of the IMF extension and the HCS features, demonstrated in Figure 9 (and even its improved versions), unsatisfactory describes the complex HCS behaviour at 1 AU, so the HCS position can not be easily calculated. We should use the real solar wind data for the analysis of the HCS inclination, which could be a possible source of the two-humped distribution of the horizontal IMF components.

The most significant investigation of the HCS inclination belongs to Lepping *et al.* (1996). They have used the variance analysis of 212 HCS crossings for 5 months in 1994-1995, and have shown the results in the picture of the distribution of longitudes of the HCS normals for the values range 225°±90° (where 225° is ortho-Parker direction). We used here their results for mapping a histogram in Figure 10a. Theoretically the distribution of normals must be centred at 225°, i.e. the HCS front must be parallel to Parker Spiral (see Figure 9). So Lepping *et al.* (1996) have expected to see the Gauss-like shape of the distribution. But obtained distribution was bimodal; the solar wind data have been showing that there is a big spread of longitude values from 135° to 315°.

The bimodal shape of the distribution in Figure 10a could be a result of a wrong prerequisite. Let's rearrange data in Figure 10a relative not to the Parker Spiral, but to the ecliptic plane (see Figure 9b). For our task we will be looking only at a degree of verticality of the HCS' front towards the ecliptic plane, so we can plot the normals' longitudes histogram not for all 360° or 180° (like Lepping *et al.* (1996) did), but merely for 90°, folding the coordinates plane in Figure 9b in four. The histogram of normals' longitudes in range 180°+90° is shown in Figure 10b, where 180° longitude is a perpendicular to the ecliptic plane (when the HCS front is parallel to the ecliptic plane).

The new distribution looks more regular and allows to conclude that the HCS normals mainly had latitudes between the perpendiculars to the ecliptic plane and to the Parker direction for the tested period. This means that the HCS is more parallel to the ecliptic plane than it was expected. Hence, the HCS inclination is hardly a cause of investigated "two-humped distribution" effect in the horizontal components of the interplanetary magnetic field.

Furthermore, if we test an input of the HCS's zeros into the IMF, we will see a surprisingly low impact of sector boundaries on the histograms of distributions of IMF components (Figure 11). Left panel in Figure 11 demonstrates distributions of three IMF components (grey filled curves) for 1977-2009 in comparison with the distributions of the same components for days of sector boundary crossings (SBC), according to the SBC list by Dr. Leif Svalgaard (http://www.leif.org/research). SBC distributions are filled with black.

Bx, By and Bz values are distributed widely on SBC days (Figure 11a and b) (maybe, this is a result of a high IMF disturbance at the heliospheric current sheet) and have no expected high peaks at zero value of the distributions. Figures 11d, e, and (f) describe the relative input (in percents) of SBC days into each span. It shows rather significant input (up to ~30%) of the HCS into the high values of Bx, By and Bz components, so the IMF increase within SBC days is additionally confirmed by this fact. At the same time, the SBC input into the IMF Bx, By and Bz distributions for near-zero values is surprisingly low (~15%).

This picture could be a result of the HCS indication problem or characterize a physical nature of observed effect. Here we are facing an additional problem: "Where are 85% of IMF zeros coming from?" The most probable explanation of this fact and all obtained results is that observed distribution of the IMF strength consists of several distributions with different statistical features.

## 3.2 A HYPOTHESIS ON THREE DISTRIBUTIONS, MAKING ONE

The in-ecliptic IMF demonstrates complex behaviour, depending on a heliocentric distance and heliolatitude. The "two-humped" shape of the Bx, By and Br histograms at 1 AU is, obviously, just a part of the whole picture, so we have to put forward hypothesis, which would be able to explain all the facts.

The analysis of the resent successful models shows that the ideas about composing of several laws of the IMF and the solar wind expansion give the best results. For example, Owens and Crooker (2006) demonstrated a good possibility of simulation of heliospheric flux as a constant background open flux with a time-varying interplanetary CME (ICME) contribution.

Let us suppose the existence of three differently directed flows in the solar wind: two flows from the middle and high latitudes in both South and North hemispheres of the Sun (dipole closed and open flux), and one quasi-radially directed flow from the near-equator latitudes. In this case we should observe a three-humped distribution as shown in Figure 12. It is based on the observed Bx distribution (see Figure 2) and schematically demonstrates the possible shape of three distributions behind the real one. Grey and white distributions (humps) schematically show the input of the IMF from high and middle latitudes of two hemispheres of the Sun, and the central one is a former Br distribution, observed in the photosphere in the ecliptic plane.

If additionally suppose that the central distribution, first of all, has relatively low height in comparison with the two flank distributions, and, then, is significantly changing with the heliocentric distance as it is shown in Figure 6, the resulting picture at 1 AU will correspond to the observed one. Therefore, active regions and the high-latitude solar wind possibly influence the IMF at the Earth orbit in higher degree as it is supposed according to the most popular solar wind models.

The next question is about the nature of the central distribution changes with a heliocentric distance.

## 3.3 MAGNETIC RECONNECTION AT THE HELIOSPHERIC CURRENT SHEET AS A POSSIBLE "IMF ZERO'S-KILLER"

As we have shown, just 15% of near-zero values of the IMF components at 1 AU could be explained by the solar wind sector structure. We suppose that residuary 85% of zeros are related to the local current sheets. Besides, zero-lines can vanish somewhere on their way from the Sun to the Earth, as a result, zero depression of the in-ecliptic IMF is observed at 0.7-3 AU.

If we look at the properties of the heliospheric current sheet, we will see prerequisites for such a vanishing. It is plasma and the IMF turbulence at the HCS, causing magnetic reconnection. Figure 13 represents behaviour of the IMF and the solar wind density at 1 AU in temporal vicinity of SBC days (OMNI2 data from January, 1964 to April, 2010). Density $n$ and the IMF averaged magnitude |B| significantly growth around zero day, and their increase is accompanied by enhanced variability of the solar wind plasma and the IMF. Standard deviations from mean of the IMF averaged magnitude |B|, its components (Bx, By, Bz) and density $n$ have similar growth of values ±1 day around the SBC days. Figure 13c shows that Bx is disturbed stronger that all the IMF components.

The density growth is a well-known feature of sector boundaries, but the IMF increase is not so well-investigated. Indeed, when we expect to find a zero-line, we suppose another behaviour of the IMF module (see, for example, Figure 5 from the paper by Lepping *et al.*, 1996). Actually, the IMF module drops for a short period of sector boundary crossing (of the order of minutes), but the IMF strength is increased in a wide time diapason around it. In the result of daily averaging this gives pictures presented in Figures 13a and c. All increases in Figures 13 are statistically significant, as the extreme points with their standard deviations are beyond the 95% confidence interval, plotted on each side of the mean value line (see Table 1).

The above-revealed conditions redound to repeating magnetic reconnection at the HCS and, on the analogy, at local current sheets inside sectors. Those local current sheets are former local separators between the groups of sunspots of opposite sign, which are transferred into the solar wind and observed inside sectors.

The HCS is known as a zone of raised turbulence (Crooker *et al.*, 2004; Roberts, Keiter, and Goldstein, 2005; Blanco *et al*., 2006; Marsch, 2006). Dynamic processes permanently take place inside the HCS, so any large-scale instabilities near the HCS may be a cause of magnetic reconnection (see, for example, Murphy *et al.*, 1993; Gosling *et al.*, 2006), which produces waves, accelerates electrons and even heats ions (Drake *et al.*, 2009).

Magnetic reconnection in space can be repeating. Figure 13 confirms the high level of turbulence around the HCS. Recently Phan *et al.* (2006) have shown existence of the 2.5 million kilometres reconnection region. Gosling *et al.* (2007) based on the multi-spacecraft measurements found direct evidence for prolonged (at least 5 hours) magnetic reconnection at a continuous X-line in the solar wind. Then Phan, Gosling, and Davis (2009) published their result on investigation of 51 HCS and concluded that "reconnection X-lines in large-scale current sheets are fundamentally extended, and not patchy and randomly distributed in space".

Magnetic reconnection at current sheets changes field structure in heliosphere. Gosling *et al.* (2007) mentioned this effect in their work. They believe that number of magnetic lines, originally connected with the Sun, is reduced by prolonged reconnection. Repeating reconnection produces discontinuities, waves and additional local X-lines around the main current sheet. Roberts, Keiter, and Goldstein (2005) confirmed this statement and found out that the HCS structure becomes more and more turbulent and complex with the distance from the Sun.

Magnetic reconnection could be one of the possible causes of complexity and multiplication of HCS. We also can suppose that the "two-humped" in-ecliptic IMF effect is weaken with distance from the Sun (see Figures 5 and 7), because current sheets become thinner and thinner under repeating reconnection. Possibly, reconnection will stop when a current sheet's thickness dwindles to some limit.

## 4. Discussion and conclusions

The observed two-humped shape of the in-ecliptic IMF distribution at 1 AU is far from the expected one according to the most popular solar wind and the IMF expansion models. There is bright depression of zero values of the in-ecliptic IMF (Bx, By or Br) at 1AU. At the same time, there is no such an effect in a shape of the radial photospheric magnetic field (Br) histogram.

1. We have found that the effect varies with time and distance:

   - The IMF histograms' shape depends on a solar cycle (the histograms fall down and spread at the solar maxima). Changes in the solar magnetic field Br distribution are most expressed, but Bx and By do not look significantly influenced by Br behaviour.

   This fact is one of the indirect confirmations of multi-impact of the solar wind streams from different solar latitudes into the in-ecliptic IMF distribution. An effect of the Bx and By histograms' "falling down and spreading" at solar maximum is a result of CME and CIR more frequently coming to the Earth orbit in that period. Br histogram practically fully loses its peak around zero value at solar maximum because of multipole magnetic field of the Sun. Solar magnetic equator is twisted and often lies practically perpendicular to the ecliptic plane.

   - The two humps disapear with high distance from the Sun (the effect is strongly pronounced just from 0.7 AU to 2-3 AU at heliolatitudes ≤40°).

2. It was shown that the sector structure is not only cause of the effect. The heliospheric current sheet gives merely 15% of IMF zeros, observed at the Earth orbit. The HCS inclination can not influence the shape of the in-ecliptic IMF distribution too. Input of the HCS into the centre of in-ecliptic IMF distribution is relatively small, so it is logically to assume that other zeros are from local separators between sunspot groups. They are observed at 1 AU as relatively thin current sheets in sectors of the certain sign.

3. Additionally, we have tested a hypothesis about possible zeros' "spilling over" from Bx and By into Bz due to possible sharp inclination of the HCS relative to the ecliptic plane. Data analysis has shown that actually the HCS' fronts mainly lie between the predicted direction of the Parker Spiral and the ecliptic plane, so the HCS is not as highly inclined as it usually supposed.

4. All obtained experimental and statistical results may be satisfactory combined and explained with no contradiction if we accept a hypothesis, that observed distribution of the in-ecliptic IMF at 1 AU is formed by three distributions. There is no changes in radial direction of the original in-ecliptic photospheric field Br, which form a central hump of the in-ecliptic IMF distribution. Streams from the middle and even high latitudes mix with Br and produce two humps of the in-ecliptic IMF distribution. So, the two flank distributions (mainly positive and negative) characterise properties of the magnetic field, coming from high/middle latitudes of two hemispheres of the Sun, and the central one is theoretically expected distribution from low latitudes of the Sun, close to the heliomagnetic equator.

5. We suggest that the shape of the central distribution is affected by nonlinear processes in space, mainly, by magnetic reconnection at zero-line (X-line) in the HCS and local current sheets.

This process can take place during reconnection both in solar corona and in the solar wind. As we can see, 1 AU is a unique distance from the Sun, where we can clearly observe effects of interaction of the solar wind streams and structures. Magnetic reconnection is brightly expressed on the way from the Sun to the Earth, where current sheets (both the HCS and local ones) are enough thick and extensive. The solar magnetic field fast weakens with distance. X-lines get thin partially because of reconnection, which leads to their splitting and surrounds them with cloud of nonlinear waves, discontinuities, magnetic holes, and low-entropy structures. Disappearance of the "two-humped" effect with a distance is an interesting fact, which may indicate first of all, significant and "non-linear" change of the IMF picture with the increasing disctance from the Sun, and, then, fast losing of solar wind conditions, favourable for a magnetic reconnection. We suppose that this process goes repeatedly in enough thick current sheets only and under conditions of rather strong magnetic fields around X-lines. Possibly, this process goes prolonged just at the distances up to 3-4 AU.

All these hypotheses are not final and will be additionally checked subsequently. Meanwhile, there is no doubt that the investigated effect is more complex that it was assumed previously. The two-humped in-ecliptic IMF distribution can not be explained by existence of the heliospheric current sheet and sector structure only.

**Acknowledgements** OMNI2, Pioneer Venus Orbiter, Helios 2, Ulysses and Voyager 1 data were taken from the official Goddard Space Flight Center OMNIweb plus website: http://omniweb.gsfc.nasa.gov We used here the SBC List by Leif Svalgaard from his official web-page: http://www.leif.org/research/sblist.txt. The data of Wilcox Solar Observatory were used for calculation of the solar wind source surface magnetic field (http://wso.stanford.edu/forms/prsyn.html). This research was supported by RFBR's grant 10-02-01063, and partially by 10-02-00277 grant.

**Figure legends**

**Figure 1** Comparison of the simulated and observed in-ecliptic IMF component $B_L$ (according to Belov, Obridko, and Shelting (2006)).

**Figure 2** Histograms of the occurrence (in the percentage of the total) of three IMF components - Bx, By, Bz in ecliptic plane at 1 AU for 1977-2009 (bars) in comparison with the distribution of the magnetic field at the source surface of the solar wind (black line behind the bars).

**Figure 3** Sunspot number variations for the tested time period, *OMNI2* data.

**Figure 4** Dependence of the IMF components histograms' shape on the solar cycle. Histograms of the occurrence of Bx, By, and Bz at 1 AU (a,b,c), and the source surface magnetic field Br (d) during solar maxima (black curves) and minima (grey filled curves).

**Figure 5** Dependence of the radial magnetic field Br histograms' view on a heliocentric distance.

**Figure 6** Shape of the Br distribution in dependence on the heliocentric distance from 0.29 AU to 1 AU, based on *Helios 2* daily data. Number of cases: 169 (a); 197 (b); 246 (c); 445 (d)

**Figure 7** Dependence of the Br IMF distribution's shape on a heliolatitude and a heliocentric distance. *Ulysses* data, 1990-2008. (a) high latitudes (b) - low latitudes. Grey curve - 1-2AU distance (11985 hours for (a) and 10372 hours of observations for (b)); thick black curve - 2-3AU distance (21017 hours for (a) and 2254 hours for (b)); thin black curve - 3-4 AU (19467 hours for (a) and 5984 hours for (b)).

**Figure 8** Dependence of the Br IMF distribution's shape on a heliolatitude in 2007. Black curve - Br at low latitudes (2773 hours), grey- and white- filled curves - Br at high latitudes (1395 and 1659 hours of *Ulysses'* observations respectively).

**Figure 9** Orthodox views on the IMF expansion and the heliospheric current sheet (after Hundhausen, 1972).

**Figure 10** Histograms of the longitudes of the current sheet normals according to Lepping et al. (1996). (a) Original view of the histogram with the centre on ortho-Parker direction (225°±90°). The distribution is bimodal. (b) Rearranged histogram of the HCS normals' longitudes from 180° to 270° (the HCS verticality test). Quasi-unimodal distribution.

**Figure 11** Input of sector boundaries into the interplanetary magnetic field pattern. (a,b,c) Histograms of the occurrence of three components of the IMF (Bx, By, Bz - GSE) at 1 AU for 1977-2009 (grey filled curves) in comparison with histograms of the same parameters for sector boundaries crossing days (black filled curves). (d,e,f) Percentage input of the sector boundaries into each span of the corresponding IMF histograms (a-c).

**Figure 12** One possible explanation of the "two-humped" shape of the in-ecliptic IMF distribution (the real Bx distribution from Figure 2 is taken for the example).

**Figure 13** Superposed epoch analysis results for solar wind parameters around the days of sector boundaries crossings at the Earth orbit. 1300 events from January 1964 to April 2010, according to the SBC list by Leif Svalgaard were considered. Significant growth of the IMF averaged

magnitude (|B|) and solar wind density (*n*), as well as their variability (standard deviations of |B|, Bx, By, Bz and *n*) is observed around zero-day.

Tables

**Table 1** Mean values, 95% confidence interval, and standard deviations at maxima for the parameters in Figure 13

|  | **\|B\|** | st.dev.\|**B**\| | st.dev.**B**x | st.dev.**B**y | st.dev.**B**z | ***n*** | st.dev.***n*** |
|---|---|---|---|---|---|---|---|
| mean | 5.44 | 1.05 | 1.67 | 1.9 | 1.81 | 5.67 | 1.76 |
| 95% confidence interval | 0.21 | 0.08 | 0.09 | 0.11 | 0.10 | 0.34 | 0.19 |
| standard deviation at maximum | 3.93 | 1.55 | 1.57 | 2.4 | 1.82 | 6.31 | 3.45 |

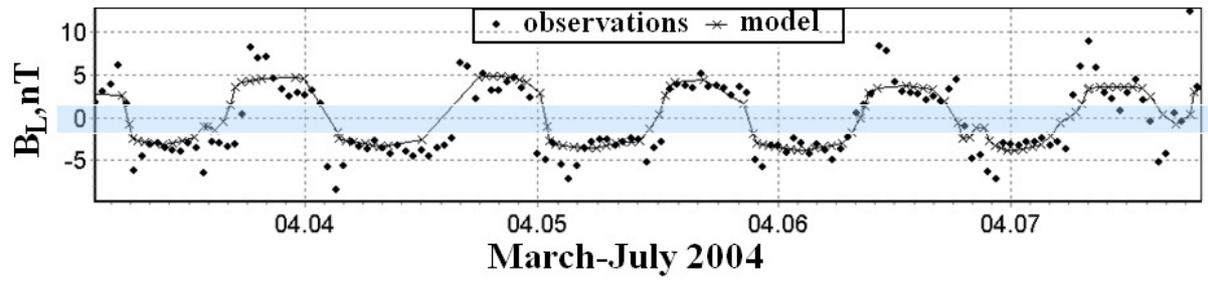

**Figure 1** Comparison of the simulated and observed in-ecliptic IMF component $B_L$ (according to Belov, Obridko, and Shelting (2006)).

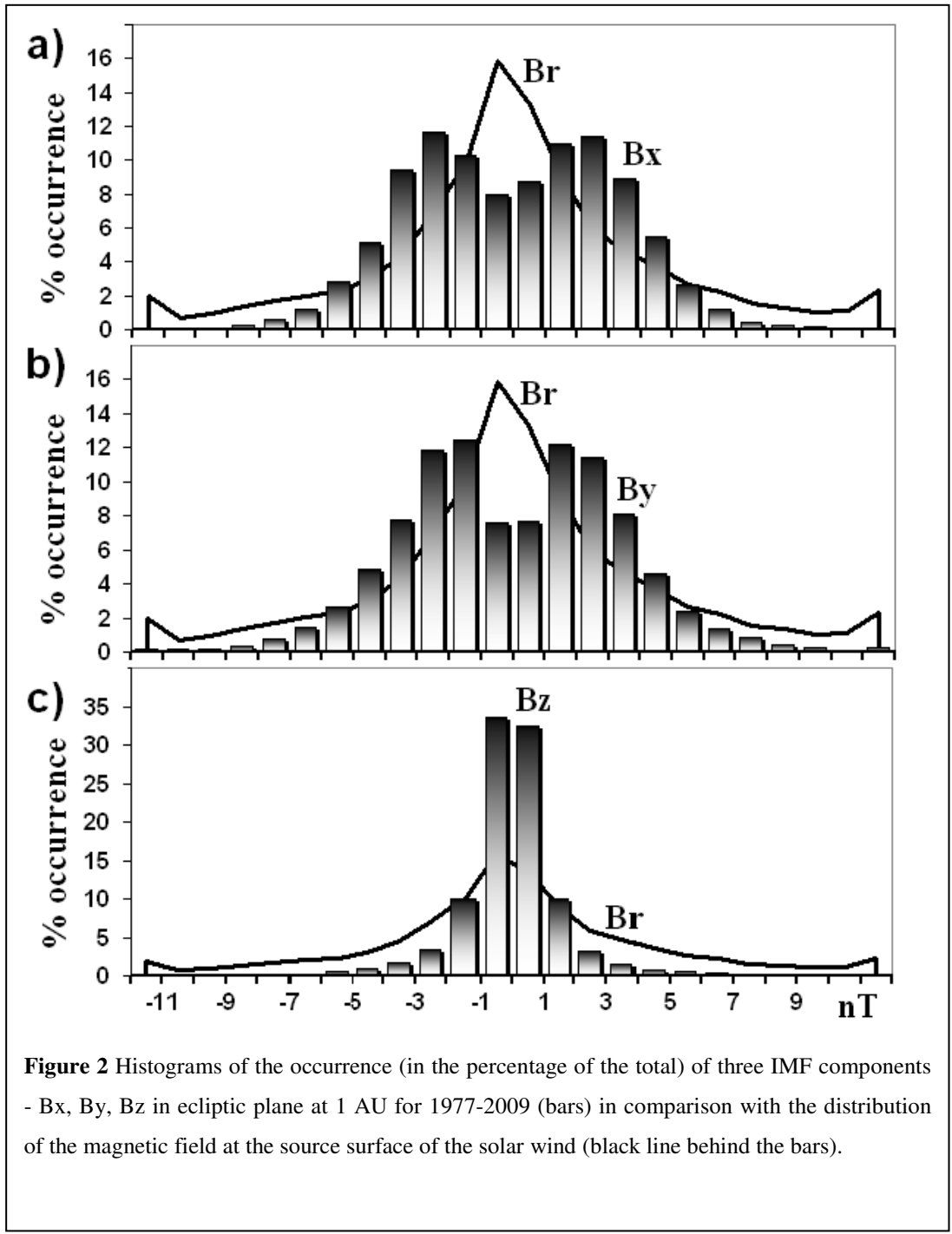

**Figure 2** Histograms of the occurrence (in the percentage of the total) of three IMF components - Bx, By, Bz in ecliptic plane at 1 AU for 1977-2009 (bars) in comparison with the distribution of the magnetic field at the source surface of the solar wind (black line behind the bars).

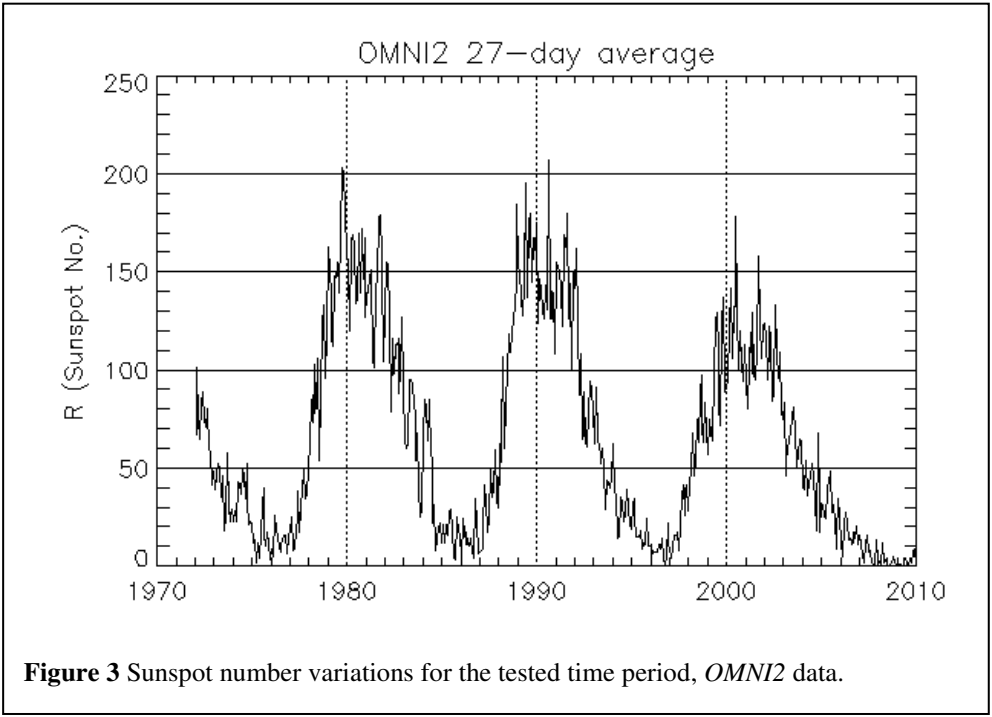

**Figure 3** Sunspot number variations for the tested time period, *OMNI2* data.

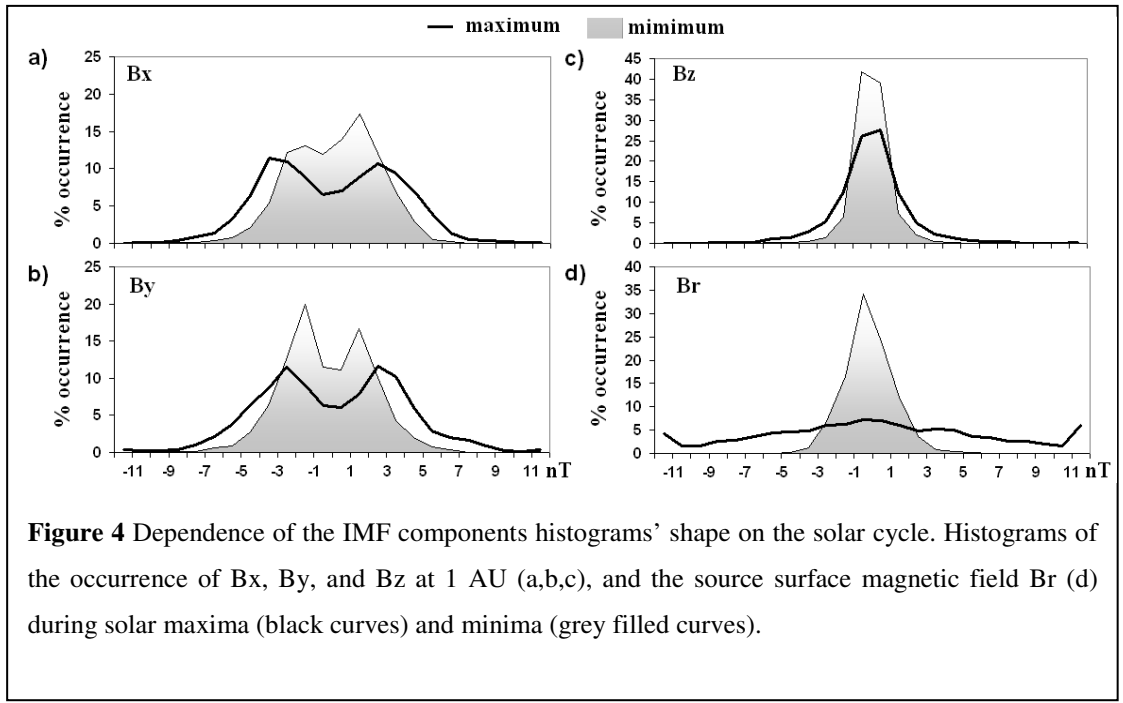

**Figure 4** Dependence of the IMF components histograms' shape on the solar cycle. Histograms of the occurrence of Bx, By, and Bz at 1 AU (a,b,c), and the source surface magnetic field Br (d) during solar maxima (black curves) and minima (grey filled curves).

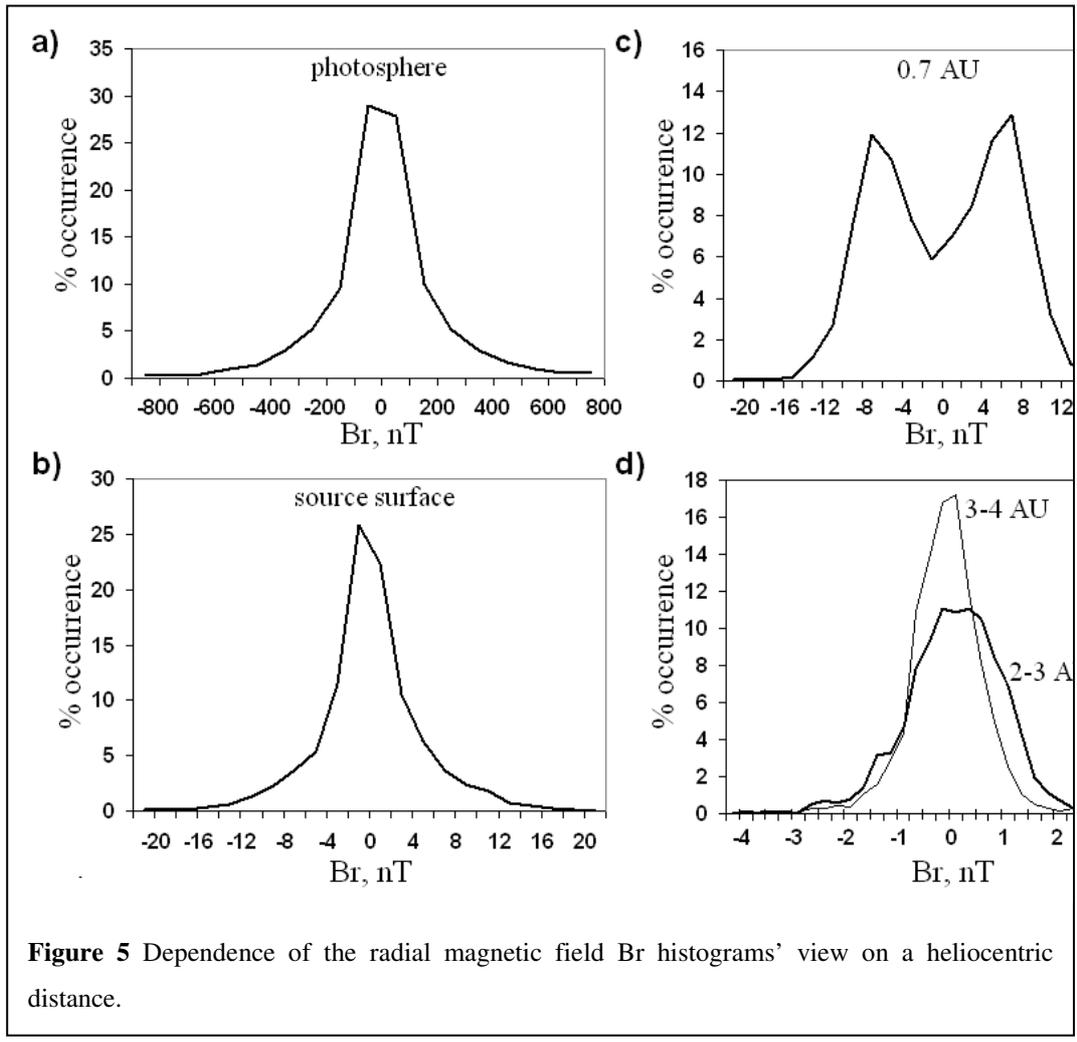

**Figure 5** Dependence of the radial magnetic field Br histograms' view on a heliocentric distance.

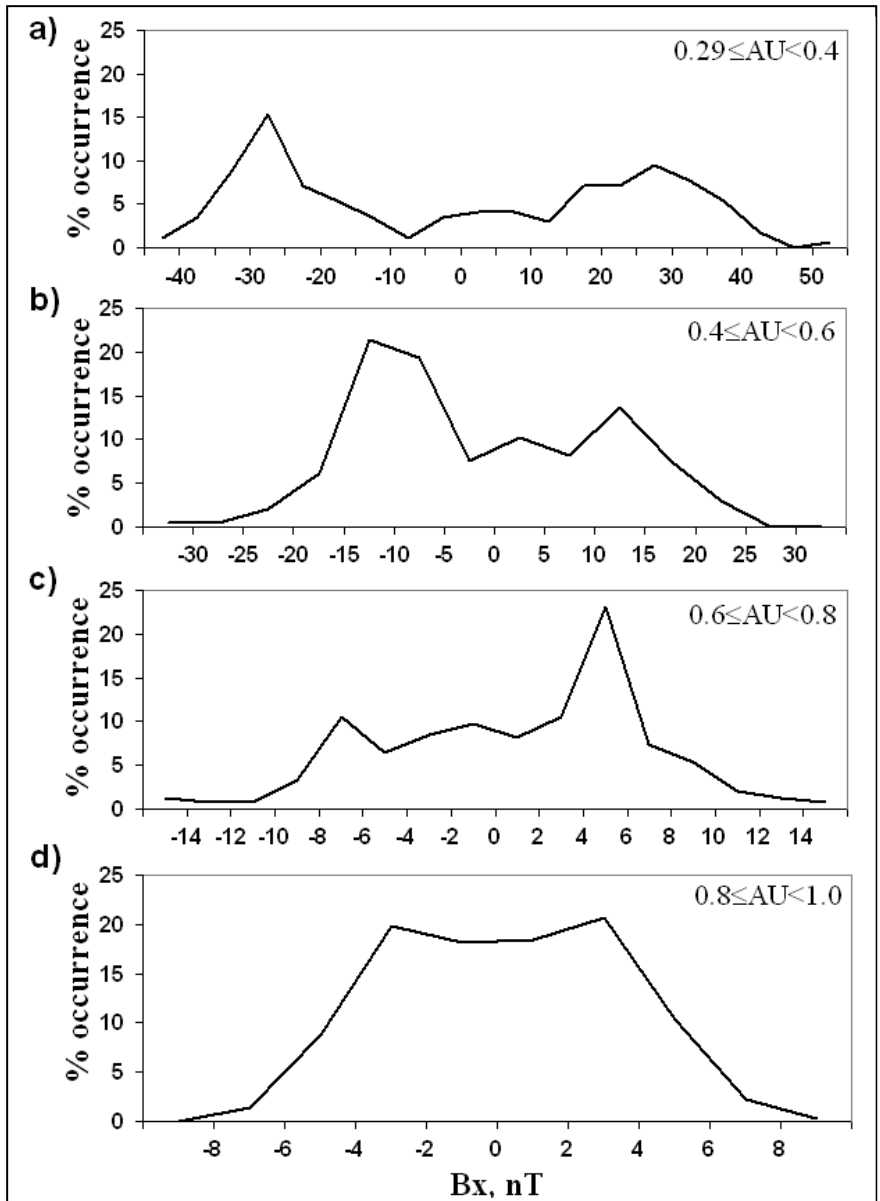

**Figure 6** Shape of the Br distribution in dependence on the heliocentric distance from 0.29 AU to 1 AU, based on *Helios 2* daily data. Number of cases: 169 (a); 197 (b); 246 (c); 445 (d)

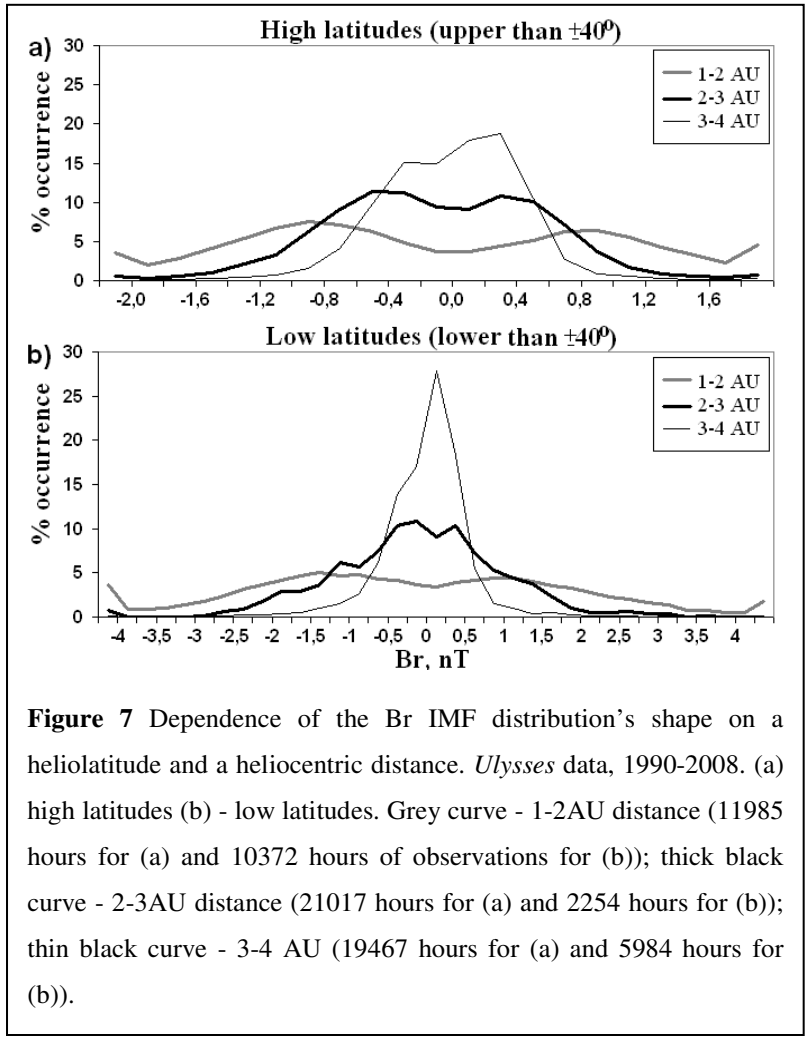

**Figure 7** Dependence of the Br IMF distribution's shape on a heliolatitude and a heliocentric distance. *Ulysses* data, 1990-2008. (a) high latitudes (b) - low latitudes. Grey curve - 1-2AU distance (11985 hours for (a) and 10372 hours of observations for (b)); thick black curve - 2-3AU distance (21017 hours for (a) and 2254 hours for (b)); thin black curve - 3-4 AU (19467 hours for (a) and 5984 hours for (b)).

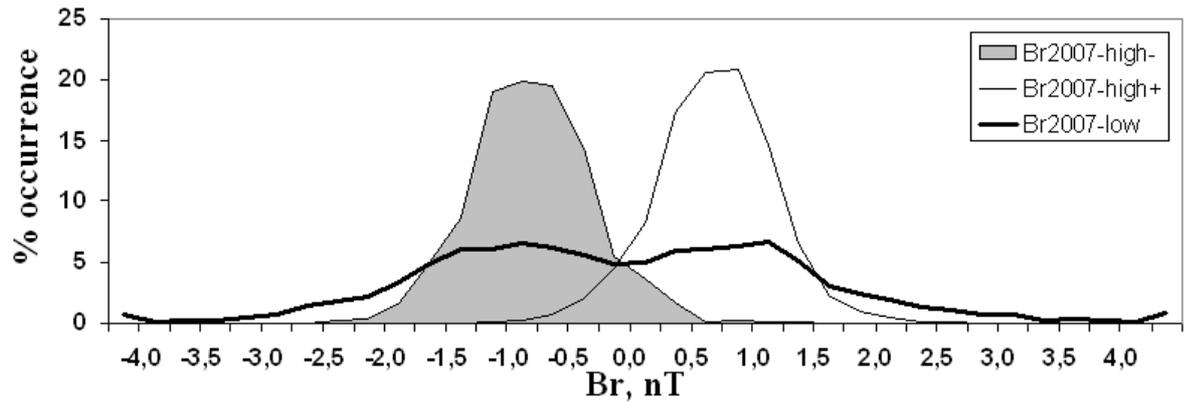

**Figure 8** Dependence of the Br IMF distribution's shape on a heliolatitude in 2007. Black curve - Br at low latitudes (2773 hours), grey- and white- filled curves - Br at high latitudes (1395 and 1659 hours of *Ulysses'* observations respectively).

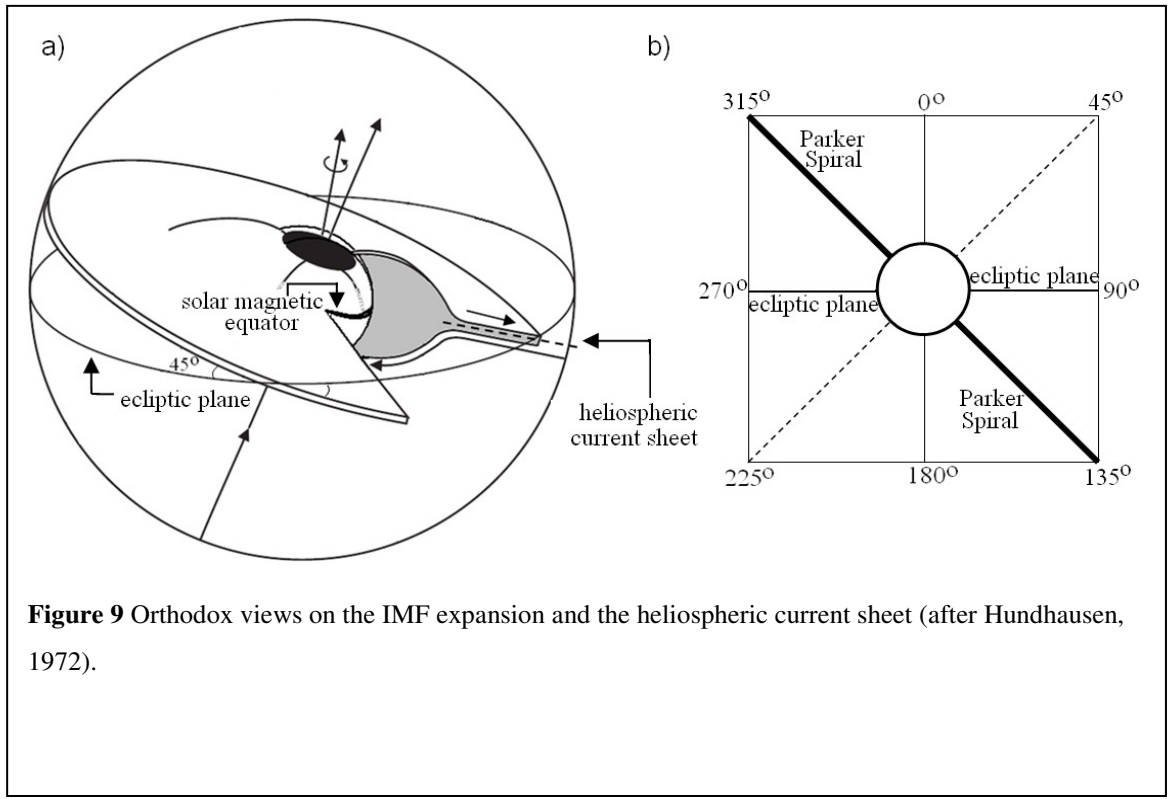

**Figure 9** Orthodox views on the IMF expansion and the heliospheric current sheet (after Hundhausen, 1972).

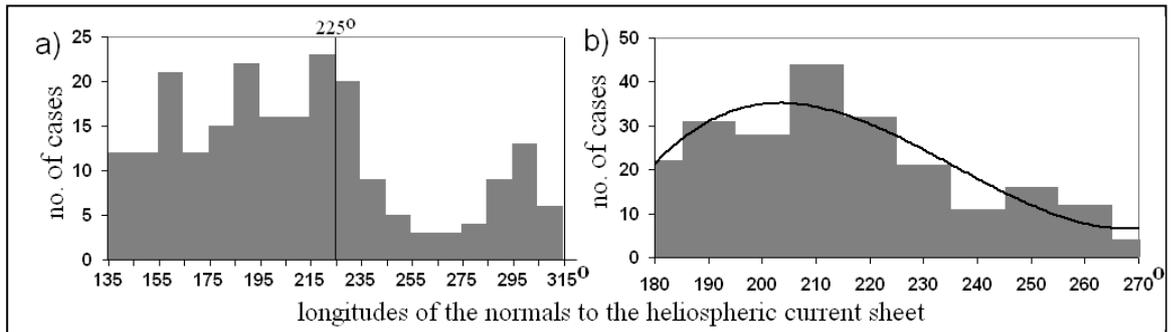

**Figure 10** Histograms of the longitudes of the current sheet normals according to Lepping et al. (1996). (a) Original view of the histogram with the centre on ortho-Parker direction (225°±90°). The distribution is bimodal. (b) Rearranged histogram of the HCS normals' longitudes from 180° to 270° (the HCS verticality test). Quasi-unimodal distribution.

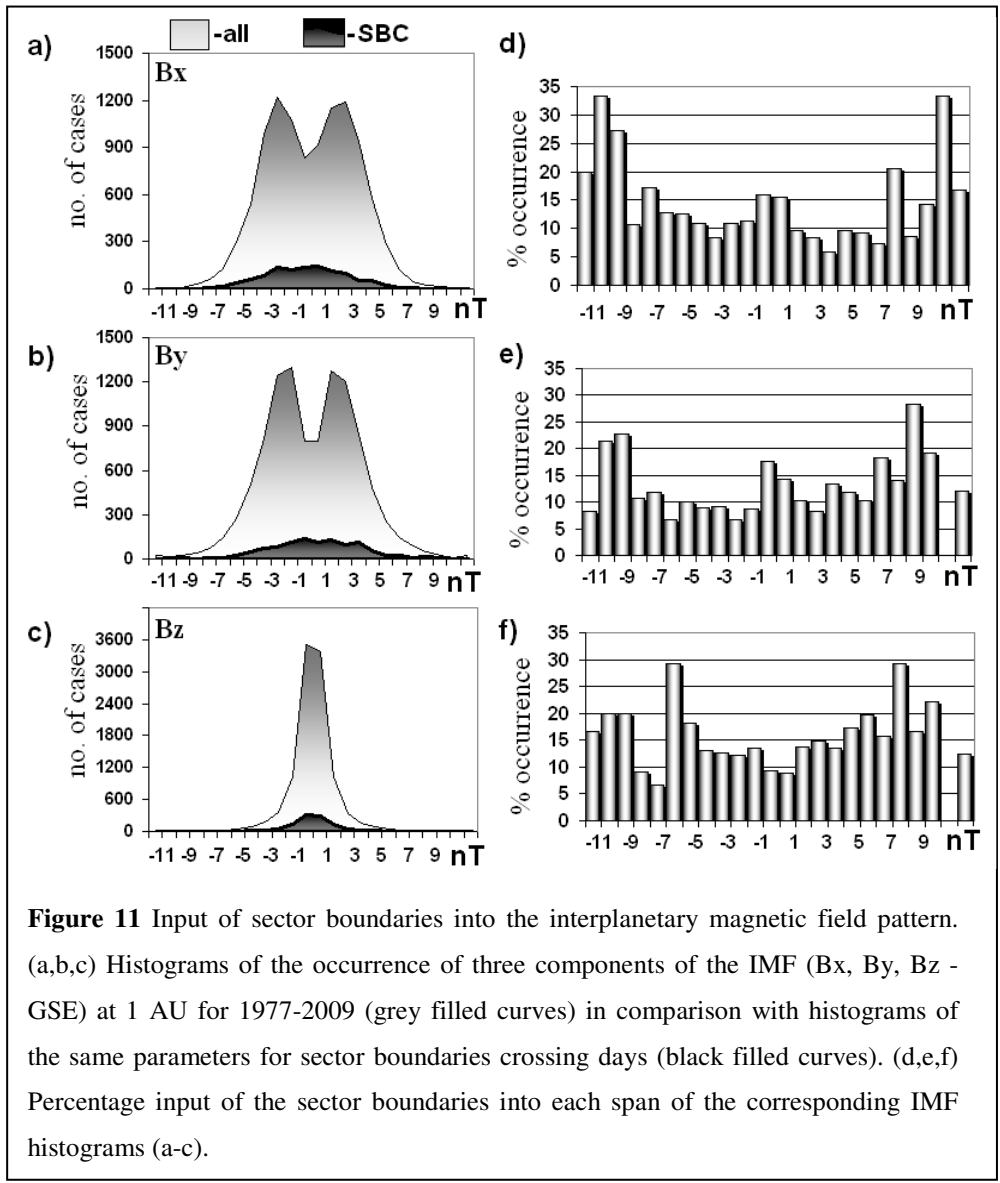

**Figure 11** Input of sector boundaries into the interplanetary magnetic field pattern. (a,b,c) Histograms of the occurrence of three components of the IMF (Bx, By, Bz - GSE) at 1 AU for 1977-2009 (grey filled curves) in comparison with histograms of the same parameters for sector boundaries crossing days (black filled curves). (d,e,f) Percentage input of the sector boundaries into each span of the corresponding IMF histograms (a-c).

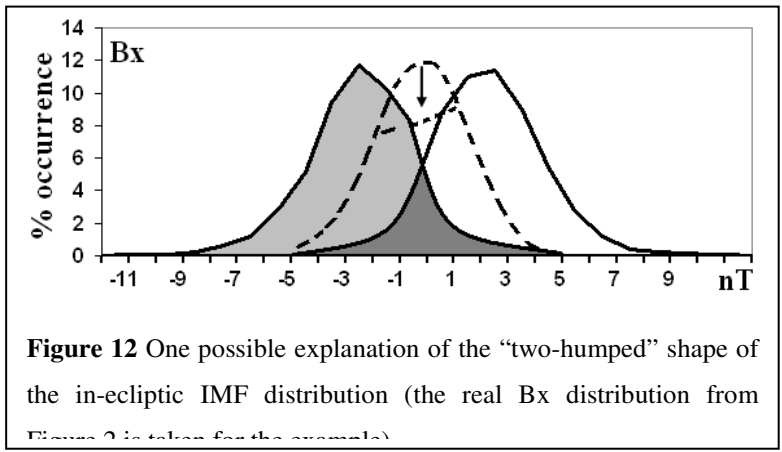

**Figure 12** One possible explanation of the "two-humped" shape of the in-ecliptic IMF distribution (the real Bx distribution from Figure 2 is taken for the example)

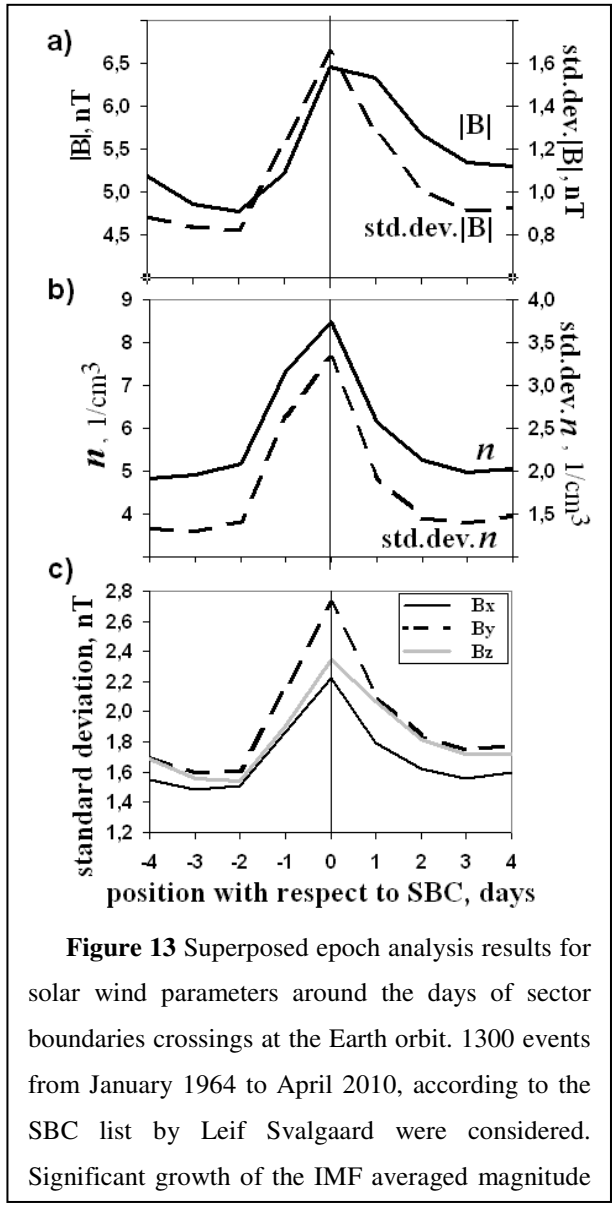

**Figure 13** Superposed epoch analysis results for solar wind parameters around the days of sector boundaries crossings at the Earth orbit. 1300 events from January 1964 to April 2010, according to the SBC list by Leif Svalgaard were considered. Significant growth of the IMF averaged magnitude